\documentclass[aps,twocolumn,preprintnumbers,amsmath,amssymb,superscriptaddress]{revtex4-1}

\usepackage{amsmath,amssymb,amsfonts,mathrsfs} 
\usepackage{amsthm}
\usepackage{CJK}
\usepackage{bm}
\usepackage{graphicx}
\usepackage{color}
\usepackage{hyperref}

\usepackage{pdfpages}
\makeatletter
\AtBeginDocument{\let\LS@rot\@undefined}
\makeatother

\def \beq {\begin{equation}}
	\def \eeq {\end{equation}}

\def \k {\mathbf{k}}

\DeclareMathOperator{\diag}{diag}

\begin{document}

\title{Topology and exceptional points of massive Dirac models \\ with generic non-Hermitian perturbations}

\author{W. B. Rui}
\affiliation{Max-Planck-Institute for Solid State Research, D-70569 Stuttgart, Germany}

\author{Y. X. Zhao}
\email[]{zhaoyx@nju.edu.cn}
\affiliation{National Laboratory of Solid State Microstructures and Department of Physics, Nanjing University, Nanjing 210093, China}
\affiliation{Collaborative Innovation Center of Advanced Microstructures, Nanjing University, Nanjing 210093, China}

\author{Andreas P.\ Schnyder}
\email[]{a.schnyder@fkf.mpg.de}
\affiliation{Max-Planck-Institute for Solid State Research, D-70569 Stuttgart, Germany}

\date{\today}

\begin{abstract}
Recently, there has been a lot of activity in the research field of topological non-Hermitian physics, partly driven by fundamental interests and partly driven by applications in photonics. However, despite these activities, a general classification and characterization of non-Hermitian
Dirac models that describe the experimental systems is missing. Here, we present a systematic investigation of massive Dirac models on periodic lattices, perturbed by general non-Hermitian terms. We find that there are three different types of non-Hermitian terms.
For each case we determine the bulk exceptional points,  the boundary modes, and the band topology. 
Our findings serve as guiding principles for the design of
applications, for example, in photonic lattices.
For instance, periodic Dirac systems with non-Hermitian mass terms  can be used as topological lasers.
Periodic Dirac systems with non-Hermitian anti-commuting terms, on the other hand,
exhibit exceptional points at the surface, whose non-trivial topology
could be utilized for optical devices.
\end{abstract}

\vspace{1.4cm}

\pacs{ }

\maketitle

The fields of non-Hermitian physics and topological materials have recently intertwined to create 
the new research direction of non-Hermitian topological phases.
As a result of the joint efforts from both fields, fascinating new discoveries
have been made, both at the fundamental level and with respect to applications~\cite{chen_exceptional_sensor_2017,bandres_topological_2018,harari_topological_2018,bahari_top_laser_science_2017,feng_single_mode_2014,hodaei_parity_timesymmetric_2014,goldzak_light_2018,weimann_topologically_2017,cerjan_experimental_2018,leykam_edge_2017,pan_photonic_2018,gong_topological_2018,lee_anomalous_2016,martinez_alvarez_non-hermitian_2018,yao_edge_2018,lee_anatomy_2018,shen_topological_2018,yao_non-hermitian_2018,kunst_biorthogonal_2018,kawabata_anomalous_2018,zhen_spawning_2015,zhou_observation_2018,xu_weyl_2017,Kawabata_Sato_arXiv_18}. 
For instance, topological exceptional points have been found both in non-Hermitian periodic lattices~\cite{lee_anomalous_2016,martinez_alvarez_non-hermitian_2018,yao_edge_2018,lee_anatomy_2018,shen_topological_2018,yao_non-hermitian_2018,kawabata_anomalous_2018,kunst_biorthogonal_2018} and 
in non-Hermitian non-periodic systems~\cite{Miri_science_review7709}. 
% in one-dimensional non-Hermitian lattices~\cite{lee_anomalous_2016,martinez_alvarez_non-hermitian_2018,yao_edge_2018,lee_anatomy_2018}
%and in non-Hermitian Chern insulators~\cite{shen_topological_2018,yao_non-hermitian_2018,kawabata_anomalous_2018,kunst_biorthogonal_2018}.
Exceptional rings and  bulk Fermi arcs have been discovered in non-Hermitian periodic lattices with semi-metallic band structures \cite{zhen_spawning_2015,zhou_observation_2018,xu_weyl_2017,papaj_nodal_2018,carlstrom_exceptional_2018,yang_nodal_2018,cerjan_effects_2018,Budich_symmetry_2019,Okugawa_Topological_2019}. 
At these exceptional points and rings, two or more eigenstates become identical and self-orthogonal, leading to a defective Hamiltonian with a nontrivial Jordan normal form~\cite{heiss_physics_2012}. 
These exceptional points %manifolds 
have many interesting applications.
For example, they 
can be used as sensors with enhanced sensitivity~\cite{chen_exceptional_sensor_2017,wirsig_PRL_14}, 
as optical omni-polarizers~\cite{christodoulides_omnipolarizer_PRL_17}, 
for the creation of chiral laser modes~\cite{peng_chiral_laser_modes_PNAS_16}, 
%and stopping of light  in coupled optical waveguides~\cite{goldzak_light_2018}.,
or for unidirectional light transmission~\cite{berini_unidirectional_transmission_NatCommun_17,lin_unidirectional_invisibility_PRL_11,regensburger_unidirectional_invisibility_Nature_12}.
Furthermore, non-Hermitian periodic lattices with   nontrivial topology  can be utilized
as topological lasers~\cite{bahari_top_laser_science_2017,harari_topological_2018,bandres_topological_2018}. Moreover, it has been shown that non-Hermitian topological Hamiltonians provide useful descriptions of strongly correlated materials
in the presence of disorder or dissipation~\cite{zyuzin_flat_2018,kozii_non-hermitian_2017,yoshida_non-hermitian_2018,lourenco_kondo_2018,lourenco_kondo_2018,nakagawa_non-hermitian_2018,kawabata_parity-time-symmetric_2018,avila_non-hermitian_2018,shen_quantum_2018,harrison_highly_2018}. This has given new insights into the Majorana physics of semiconductor-superconductor
nanowires~\cite{avila_non-hermitian_2018} and into the quantum oscillations of SmB$_6$~\cite{shen_quantum_2018,harrison_highly_2018}.

Despite these recent activities, a general framework for the study and classification of non-Hermitian Dirac models that describe
the aforementioned experimental systems is lacking. In particular, a general classification of exceptional points and 
topological surface states in non-Hermitian systems is missing, 
%the formulation of bulk topological invariants, the associated bulk-boundary correspondence, and the role of boundary conditions are still unclear for non-Hermitian topological phases, 
although various attempts have been made with partial success for certain special cases~\cite{yao_edge_2018,lee_anomalous_2016,shen_topological_2018,yao_non-hermitian_2018,martinez_alvarez_non-hermitian_2018,lee_anatomy_2018,yao_edge_2018,yao_non-hermitian_2018,kunst_biorthogonal_2018,kawabata_anomalous_2018,kawabata_parity-time-symmetric_2018,gong_topological_2018,zhou_lee_arXiv_18}. 
Since most non-Hermitian experimental systems can be faithfully described by Dirac models with small non-Hermitian perturbations~\cite{bandres_topological_2018,harari_topological_2018,feng_single_mode_2014,hodaei_parity_timesymmetric_2014,goldzak_light_2018,parto_edge-mode_2018,yao_electrically_2018,zhao_topological_2018,st-jean_lasing_2017,weimann_topologically_2017,ruter_observation_2010,feng_non-hermitian_2017},
a systematic investigation of general non-Hermitian perturbations of  Dirac Hamiltonians
would be particularly valuable. 
This would be not only of fundamental interest, but could also inform the design of new applications.

In this Rapid Communication, we present  %a systematic investigation by considering $d$-dimensional 
a systematic investigation of massive Dirac Hamiltonians on periodic lattices,
perturbed by small non-Hermitian terms. 
We show that these non-Hermitian Dirac Hamiltonians can be either intrinsically or superficially non-Hermitian, depending on whether the non-Hermiticity can be removed
by a similarity transformation. % with open or periodic boundary conditions. 
According to the Clifford algebra, general non-Hermitian terms can be categorized into three different types:
(i) non-Hermitian terms that anti-commute with the whole Dirac Hamiltonian, (ii) kinetic non-Hermitian terms, and (iii) non-Hermitian mass terms.  
Remarkably, we find a two-fold duality
for the first two types of non-Hermitian perturbations: 
Dirac models perturbed by type-(i) terms are superficially non-Hermitian 
with periodic boundary conditions (PBCs),
but intrinsically non-Hermitian with open boundary conditions (OBCs). 
Vice versa, Dirac models with type-(ii) terms are intrinsically non-Hermitian with PBCs, but
superficially non-Hermitian with OBCs.
Interestingly, 
for type-(i) and type-(ii) terms the non-Hermiticity leads to $(d-2)$-dimensional exceptional spheres
in the surface and bulk band structures, respectively. 
Type-(iii) terms, on the other hand, induce intrinsic non-Hermiticity both for OBCs and PBCs, but
with a purely real surface-state spectrum and no exceptional spheres.

\textit{Intrinsic versus\ superficial non-Hermiticity.---}
We begin by discussing some general properties of non-Hermitian physics.
First, we recall that in Hermitian physics only unitary transformations of the Hamiltonian
are considered, because only these preserve the  reality of the expectation values.
In non-Hermitian physics, however, the Hamiltonian can be similarity transformed,
$H\rightarrow V^{-1}HV$, by any invertible matrix $V$, which is not necessarily unitary but is required to be local. 
For this reason, a large class of non-Hermitian 
 Hamiltonians $H$  can be converted into Hermitian ones by non-unitary similarity transformations, i.e.,
\begin{equation} \label{similarity-trans} 
V^{-1}HV=H',\quad H'^\dagger=H'. 
\end{equation}
Using this observation, we call Hamiltonians whose non-Hermiticity can or cannot be removed by the above transformation
as superficially or intrinsically non-Hermitian, respectively. 

For non-interacting local lattice models, which is our main focus here, $H$ is a quadratic form, whose 
entries are specified as  $H_{(\mathbf{r}\alpha),(\mathbf{r}'\alpha')}$ with $\mathbf{r}$ the positions of the unit cells and $\alpha$ a label for internal degrees of freedom. Correspondingly, the similarity transformation has matrix elements $V_{(\mathbf{r}\alpha),(\mathbf{r}'\alpha')}$.
By the locality condition, the matrix elements $H_{(\mathbf{r}\alpha),(\mathbf{r}'\alpha')}$ and $V_{(\mathbf{r}\alpha),(\mathbf{r}'\alpha')}$
are required to tend to zero sufficiently fast as $|\mathbf{r}-\mathbf{r}'|\rightarrow \infty$.
If $H$ can be converted into a Hermitian Hamiltonian by a local transformation $V$, its eigenvalues are necessarily real.
Conversely, any local lattice Hamiltonian with a real spectrum is either  entirely Hermitian or superficially non-Hermitian
\footnote{We note that, for instance, space-time-inversion symmetric (i.e., $\mathcal{PT}$ symmetric) spinless lattice models have always real spectra and are 
therefore only superficially non-Hermitian.}.

A characteristic feature of non-Hermitian lattice models is the existence of 
exceptional points in parameter space, where two or multiple eigenvectors
become identical, leading to a non-diagonalizable Hamiltonian. However,
it is important to note that such exceptional points
are not dense in parameter space. That is, there  exist arbitrarily small perturbations
which remove the exceptional points, rendering the Hamiltonian diagonalizable
\footnote{It should be noted, however, that in  periodic Hamiltonians exceptional points can be stable,
in the sense that small perturbations cannot remove them from the BZ.}. 
One such perturbation relevant for lattice models are the boundary conditions~\cite{kunst_biorthogonal_2018,kunst_transfer_matrix_2018}, 
which modify the hopping amplitudes between opposite boundaries. 
For a general classification of non-Hermitian Hamiltonians, it is therefore essential to
distinguish between different types of boundary conditions, in particular OBCs and PBCs.
With PBCs and assuming translation symmetry, we can perform a Fourier transformation 
of Eq.~\eqref{similarity-trans} to obtain
 $H'(\k)= V^{-1}(\k) H (\k) V (\k)$.
Here, $ V (\k)$ is assumed to be local in momentum space. 
It is worth noting that the locality in momentum space is essentially different from that in real space. 
Generically, %$V_{(\mathbf{r}\alpha),(\mathbf{r}'\alpha')}$ has no translation symmetry, and 
the Fourier transform of $ {V}(\k)$,  $V_{\mathbf{r},\mathbf{r}'}=\sum_{\mathbf{k}} {V}(\k)e^{i\mathbf{k}\cdot (\mathbf{r}-\mathbf{r}')}$, is not local in general.

\textit{Non-Hermitian Dirac Hamiltonians.---}
We now apply the above concepts to non-Hermitian Dirac models of the form $H=H_0+\lambda U$, 
where $H_0$ is a Hermitian Dirac Hamiltonian with mass $M$, and $U$ a non-Hermitian perturbation
with $\lambda \ll M$. Assuming PBCs in all directions, we consider 
the following Hermitian Dirac Hamiltonian on the $d$-dimensional cubic lattice 
\begin{equation}\label{general-dirac-momentum}
{H}_\text{0}(\mathbf{k})=\sum_{i=1}^{d} \sin k_i\Gamma_i+(M-\sum_{i=1}^{d}\cos k_i)\Gamma_{d+1},
\end{equation}
where $\Gamma_\mu$ denote  the gamma matrices that satisfy $\{\Gamma_\mu,\Gamma_\nu\}=2\delta_{\mu\nu}$ and $M$ is the real mass parameter. 
With OBCs in the $j$th direction and PBCs in all other directions, the Hamiltonian reads
\begin{eqnarray}\label{general-dirac-real}
H_\text{0}(\tilde{\mathbf{k}})
&=&
\frac{1}{2i}(\widehat{S}-\widehat{S}^\dagger)\otimes \Gamma_j-\frac{1}{2}(\widehat{S}+\widehat{S}^\dagger)\otimes \Gamma_{d+1}
\\
&&+\mathbb{I}_{N_j}\otimes(\sum_{i\neq j} \sin k_i\Gamma_i+(M-\sum_{i\neq j}\cos k_i)\Gamma_{d+1}),
\nonumber
\end{eqnarray}
where $\tilde{\mathbf{k}}$ denotes the vector of all momenta except $k_j$,
$\widehat{S}_{ij}=\delta_{i,j+1}$ is the right-translational operator,
and $N_j$ stands for the number of layers in the $j$th direction. %, and $\mathbb{I}_{N_j}$ is the $N_j \times N_j$ identity matrix.
From the above two equations it is now clear that, according to the Clifford algebra, 
there exist only the three types of non-Hermitian perturbations discussed above. 
We will now study these individually.

%%%%%%%%%%%%%%%%%%%%%%%%%%%%
%%%%%%%%%%%%%%%%%%%%%%%%%%%%
%%%%%%%%%%%%%%%%%%%%%%%%%%%%
\begin{figure}[t]
\includegraphics[width=1.02\linewidth]{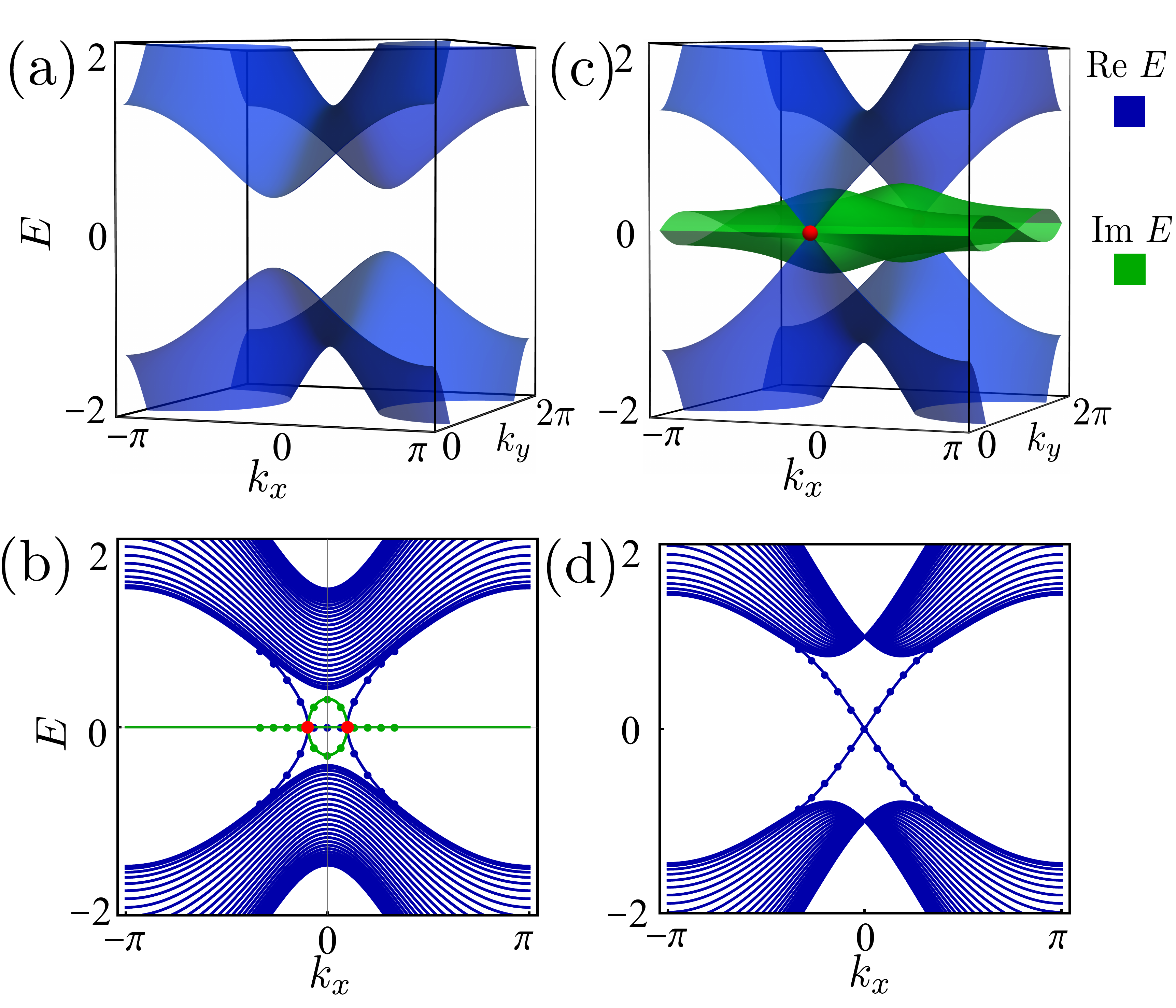} 
\caption{(a),(b) Energy spectra of the two-dimensional to\-po\-logical insulator $H_{\text{TI,0}}$
perturbed by the type-(i) non-Hermitian term $i \lambda \Gamma_4$ with
periodic and open boundary conditions, respectively. Here, we set
$\lambda=0.3$ and $M=1.5$.
(c),(d) 
Energy spectra of $H_{\text{TI,0}}$ perturbed by the type-(ii) non-Hermitian term $i \lambda \Gamma_2$
with periodic and open boundary conditions, respectively. Here, we set $\lambda=0.5$ and $M=1.5$.
Solid and dotted lines  represent bulk and surface states, respectively. The real and imaginary parts
of the eigenvalues are indicated in blue and green.
Red points represent exceptional points. \label{TI} \label{mFig1}
	}
\end{figure}
%%%%%%%%%%%%%%%%%%%%%%%%%%%%
%%%%%%%%%%%%%%%%%%%%%%%%%%%%
%%%%%%%%%%%%%%%%%%%%%%%%%%%%

%\textbf{Non-Hermitian terms of type (i).}
\textit{Non-Hermitian anti-commuting terms.---}
We start with non-Hermitian terms of type (i), i.e., terms that anti-commute with the Dirac Hamiltonian $H_0$. Such non-Hermitian terms are possible 
for all Altland-Zirnbauer classes with chiral symmetry~\cite{altlandZirnbauerPRB10,chiu_classification_2016}, 
in which case they are given by the chiral operator $\Gamma$.
With PBCs the Hamiltonian perturbed by these type-(i) terms is expressed as
\begin{equation} \label{PBC_typ_eins}
 H (\k)=  H_0(\k)+i\lambda\Gamma,
\end{equation}
where $\Gamma$ is an additional gamma matrix with $\{\Gamma, {H}_0(\k)\}=0$
and $\lambda$   a real parameter. 
The spectrum of  $H ( {\bf k} ) $ is given by $E(\k)=\pm \sqrt{d^2(\k)-\lambda^2}$ with $d^2(\k)\mathbb{I}={H}_0^2(\k)$, which is completely real for all $\k$, provided that $|\lambda|$ is smaller than the energy gap of ${H}_0$. 
Thus, Hamiltonian~\eqref{PBC_typ_eins} with $\lambda \ll M$ is only superficially non-Hermitian and we can remove the non-Hermitian term
by a similarity transformation.
The corresponding transformation matrix $V ( {\bf k} )$ can be derived systematically by noticing that the flattened Hamiltonian $\tilde{{H}}_0(\k)={H}_0(\k)/d(\k)$ and $\Gamma$ form a Clifford algebra and, thus, $i[\tilde{H}_0(\k),\Gamma]/4$ generates rotations of the plane spanned by $\tilde{H}_0(\k)$ and $\Gamma$. 
Hence, the explicit expression of the transformation matrix is $ V (\k)=\exp[- \tfrac{i}{2} \tilde{H}_0(\k)\Gamma \eta(\k)]$, with $e^{\eta(\k)}=\sqrt{(d(\k)+\lambda)/(d(\k)-\lambda)}$. 
From Eq.~\eqref{similarity-trans} it follows that the transformed Hamiltonian
is ${H}'(\k) = \sqrt{1-\lambda^2/d^2(\k)} {H}_0(\k)$, which is manifestly Hermitian for $\lambda \ll M$.

With OBCs, on the other hand, type-(i) perturbations lead to intrinsic non-Hermiticity, provided the Dirac Hamiltonian $H_0$ is
in the topological phase. This is because the topological boundary modes acquire complex spectra  due to the non-Hermitian term $i \lambda \Gamma$,
even for infinitesimally small $\lambda$. 
To see this, we first observe that for any eigenstate $\psi_0$ of $H_0$ with energy $E_0$,  $ \Gamma\psi_0$  is also an eigenstate of $H_0$, but with opposite energy $-E_0$. Applying chiral perturbation theory to $H = H_0 + i \lambda \Gamma$, we find that since $i \lambda \Gamma$ scatters
$\psi_0$ into $\Gamma \psi_0$, eigenstates of $H$ can be expressed as superpositions of  $\psi_0$ and $\Gamma\psi_0$.
Explicitly, we find that the eigenstates of $H$ are $\psi_\pm=\psi_0+c_{\pm}\Gamma\psi_0$, with  $c_{\pm}=iE_0/\lambda \pm \sqrt{1-E_0^2/\lambda^2}$
and energy  $E_{\pm}=\pm \sqrt{E_0^2-\lambda^2}$.
This analysis holds in particular also for the topological boundary modes of $H_0$, which are massless Dirac fermions with linear dispersions.
Consequently, even for arbitrarily small $\lambda$, there exists a segment in the spectrum of $H$ around $E=0$ with purely imaginary eigenergies.

To make this more explicit, we can derive a low-energy effective theory for the boundary modes, by projecting the bulk Hamiltonian onto the boundary space.
Generically, the boundary theory is of the form  ${H}_b(\tilde{\k})= \sum_{ i \ne j}  {k}_i\gamma^i+i\lambda\gamma$, where the first term describes the boundary massless Dirac fermions of ${H}_0$. The matrices $\gamma^i$ and $\gamma$ are the projections of $\Gamma_i$ and $\Gamma$, respectively, onto the boundary space, and satisfy $\{\gamma^i,\gamma\}=0$. With this, we find that the boundary spectrum is $E_b =\pm\sqrt{|\tilde{\k}|^2-\lambda^2}$, 
and that there exists a $(d-2)$-dimensional exceptional sphere of radius $| \tilde{ \k} | = \lambda$ in the boundary Brillouin zone, which separates eigenstates with purely real and purely complex energies from each other.

As an aside, we remark that even  arbitrarily large  non-Hermitian terms $i\lambda\Gamma$ cannot remove the topological surface state. 
The reason for this is that $i \lambda \Gamma$ is a chiral operator, which acts only within a unit cell and does not couple different sites. 
In other words, the expectation value of the position operator $X_i$ is independent of $\lambda$, i.e., $\frac{d}{d\lambda}\langle\psi^\alpha_\lambda|X_i|\psi^\beta_\lambda\rangle=0$ with $\langle\psi^\alpha_\lambda|$ and $|\psi^{\beta}_\lambda\rangle$ the left and right eigenstates of $H$, respectively. 

Let us now illustrate the above general considerations by considering as an example,
$
{H}_{\text{TI}}(\mathbf{k})=\sin k_x\Gamma_1+\sin k_y\Gamma_2
+(M-\cos k_x-\cos k_y)\Gamma_3+i\lambda\Gamma_4
$, 
with $\Gamma_j$ the $4\times 4$ Dirac gamma matrices,
which describes  a topological superconductor in class DIII or a topological insulator in class AII~\cite{chiu_classification_2016}. 
The energy spectra of ${H}_{\text{TI}}$ with periodic and open boundary conditions are 
shown in Figs.~\ref{mFig1}(a) and~\ref{mFig1}(b), respectively [see Supplemental Material (SM)~\cite{Supp} for details]. 
We observe that the bulk spectrum is purely real, while the surface spectrum is complex with two exceptional points
of second order located at $k_x=\pm \arcsin |\lambda|$.

%%%%%%%%%%%%%%%%%%%%%%%%%%%%
%%%%%%%%%%%%%%%%%%%%%%%%%%%%
%%%%%%%%%%%%%%%%%%%%%%%%%%%%

%%%%%%%%%%%%%%%%%%%%%%%%%%%%
%%%%%%%%%%%%%%%%%%%%%%%%%%%%
%%%%%%%%%%%%%%%%%%%%%%%%%%%%

\textit{Non-Hermitian kinetic terms.---}
We proceed by  considering non-Hermitian kinetic terms [i.e., terms of type (ii)] added to $H_0$ with PBCs.
The effects of these non-Hermitian terms can be most clearly seen by studying the continuous version of Eq.~\eqref{general-dirac-momentum}, namely 
\begin{equation} \label{ham_type_ii_perturb}
H(\mathbf{k})
= H_0 (\mathbf{k}) + i \lambda \Gamma_j
=\sum_{i=1}^d k_i \Gamma_i+m\Gamma_{d+1}+i\lambda \Gamma_j,
\end{equation}
with $1$$\le$$j$$\le$$d$ and $\lambda$ real. The energy spectrum of $H(\mathbf{k})$, $E (\mathbf{k})=\pm\sqrt{\sum_{i\neq j} k_i^2+ (k_j+i\lambda)^2+m^2}$,
is complex and exhibits exceptional points on the $(d-2)$-dimensional sphere $\sum_{i\neq j} k_i^2 + m^2=\lambda^2$ within the $k_j=0$ plane.
Hence, $H ( {\bf k} )$ with PBCs is intrinsically non-Hermitian. 

To study the case of OBCs we consider $H ( \k )$ in a slab geometry with the surface perpendicular to the $j$th direction.  The energy spectrum in this geometry
is obtained from ${H}(\tilde{\mathbf{k}},-i\partial_j)$, i.e., by replacing $k_j$ by $-i\partial_j$ in Eq.~\eqref{ham_type_ii_perturb}. Then, it is obvious that 
the non-Hermitian term $i \lambda \Gamma_j$  can be removed by the similarity transformation $V= e^{\lambda x_j}$. 
That is, $e^{-\lambda x_j}H(\tilde{\mathbf{k}},-i\partial_j)e^{\lambda x_j}= {H}_0(\tilde{\mathbf{k}},-i\partial_j)$, which is manifestly Hermitian.
Accordingly, $H(\tilde{\mathbf{k}},-i\partial_j)$ has a real spectrum and its eigenstates are related to those of  $H_0 (\tilde{\mathbf{k}},-i\partial_j)$ by $\psi(x_j,\tilde{\mathbf{k}})=e^{\lambda x_j}\psi_0(x_j,\tilde{\mathbf{k}})$. 
We conclude that continuous Dirac models perturbed by non-Hermitian kinetic terms are superficially non-Hermitian with OBCs, but
intrinsically non-Hermitian with PBCs.
The same holds true  for lattice Dirac models.

To exemplify this, we consider the lattice Dirac model of Eq.~(\ref{general-dirac-real}) perturbed by the non-Hermitian term $i \lambda \Gamma_j$, i.e.,
$H(\tilde{\mathbf{k}}) = H_0 ( \tilde{\mathbf{k}} )+\mathbb{I}\otimes i \lambda \Gamma_j$.  
This Hamiltonian can be transformed to (see SM~\cite{Supp} for details)  
\begin{multline}\label{Hprime-kinetic}
H' (\tilde{\mathbf{k}}) =\frac{1}{2i}(\widehat{S}-\widehat{S}^\dagger)\otimes \Gamma_j-\frac{1}{2}(\widehat{S}+\widehat{S}^\dagger)\otimes \Gamma_{d+1}\\
+\mathbb{I}\otimes(\sum_{i\neq j} \sin k_i\Gamma_i+\sqrt{M_k^2-\lambda^2}\Gamma_{d+1}),
\end{multline}
by the similarity transformation
$V=\diag\{1,\alpha,\cdots,\alpha^{N_j-1}\}\otimes\left[(1+\alpha) \mathbb{I} +i(1-\alpha)\Gamma_j\Gamma_{d+1}\right].$ Here $\alpha=\sqrt{(M_k-\lambda)/(M_k+\lambda)}$ and $M_k=M-\sum_{i\neq j}\cos k_i$.   Equation~\eqref{Hprime-kinetic} is  manifestly Hermitian for $\lambda \ll M$. 
%if $M^2_k\ge \lambda$ for all $\tilde{\mathbf{k}}$. When $d-1<M<d$, the condition is satisfied for $\lambda\le 1$. 
As a concrete example, we set in Eq.~\eqref{Hprime-kinetic} $d=2$ with $\Gamma_i$ the Dirac gamma matrices, which describes
a two-dimensional topological insulator. The energy spectra for this case with periodic and open boundary conditions are shown in 
Figs.~\ref{mFig1}(c) and~\ref{mFig1}(d), respectively.

%%%%%%%%%%%%%%%%%%%%%%%%%%%%%%%%%%
\begin{figure}[t]
\includegraphics[width=1.02\linewidth]{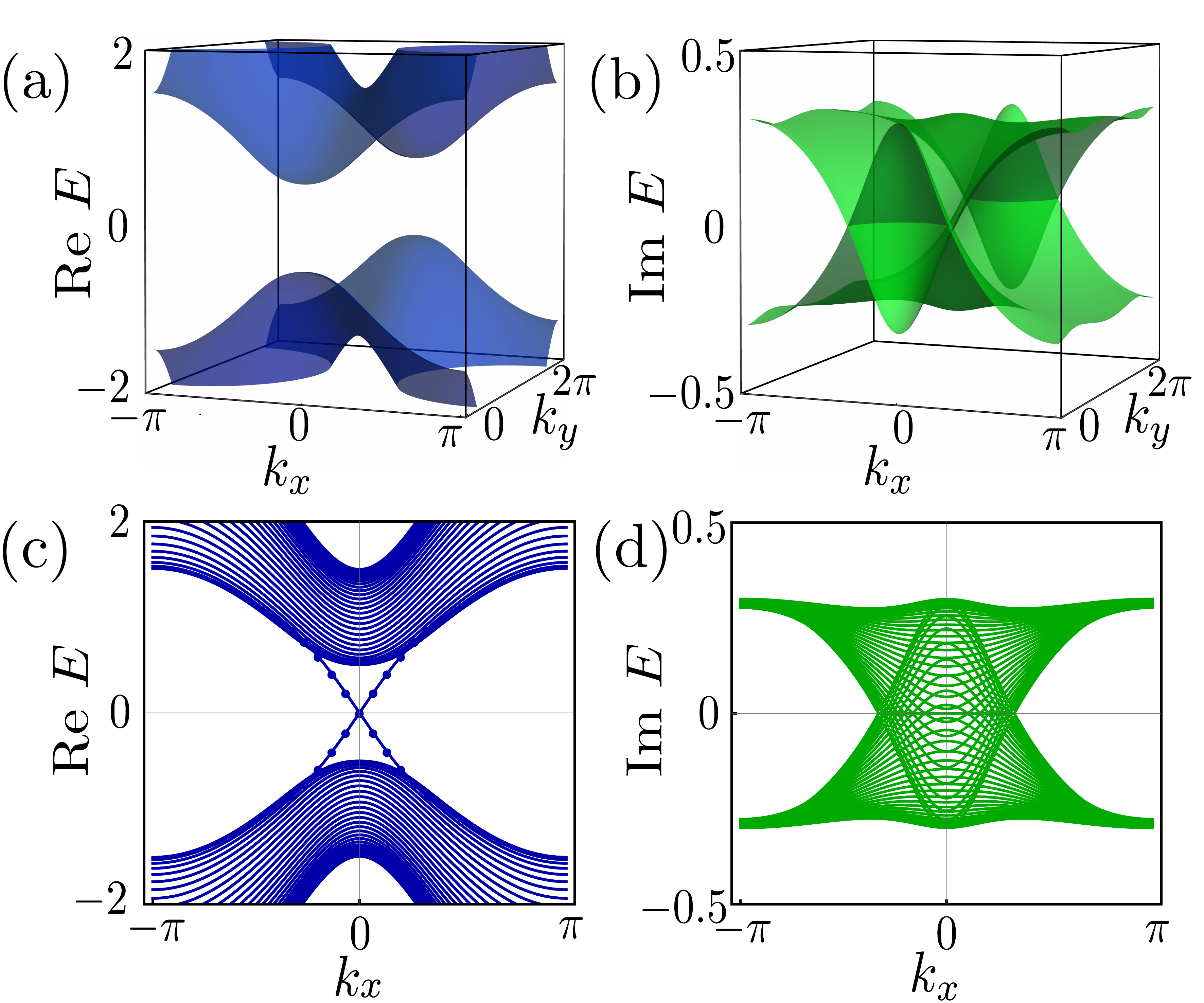} 
\caption{(a),(c) Real and (b),(d) imaginary parts of the energy spectra of the two-dimensional topological insulator $H_{\text{TI,0}}$ perturbed
by the non-Hermitian mass term    $i \lambda \Gamma_3$
with periodic and open boundary conditions, respectively. 
The parameters are chosen as $M=1.5$ and $\lambda=0.3$.
Solid and dotted lines represent bulk and surface states, respectively. 
		\label{TI2}  \label{mFig2} }
\end{figure}
%%%%%%%%%%%%%%%%%%%%%%%%%%%%%%%%%%

\textit{Non-Hermitian mass terms.---}
Finally, we examine the effects of non-Hermitian mass terms, i.e., non-Hermitian terms of type (iii). For that purpose, we add
$i\lambda\Gamma_{d+1}$ to Eqs.~\eqref{general-dirac-momentum} or \eqref{general-dirac-real}, which is equivalent to assuming that
the mass $M$ is complex. Hence, the energy spectrum is always complex independent of the boundary conditions (see SM~\cite{Supp}). Thus,
massive Dirac models perturbed by non-Hermitian mass terms are intrinsically non-Hermitian, both for open and periodic boundary conditions. 
Furthermore, we find that there are no exceptional points, not in the bulk and not in the surface band structure. 
Indeed, remarkably, topological boundary modes are unaffected by type-(iii) perturbations and keep their purely real energy spectra.

To demonstrate this explicitly, we solve for the boundary modes of Eq.~\eqref{general-dirac-real} perturbed by $i\lambda \Gamma_{d+1}$. That is, 
we solve $( H_\text{0}(\tilde{\mathbf{k}})+\mathbb{I}_{N_j}\otimes i\lambda \Gamma_{d+1} )   |\psi_{\tilde{\mathbf{k}}}\rangle = E  |\psi_{\tilde{\mathbf{k}}}\rangle$
with $|\psi_{\tilde{\mathbf{k}}}\rangle$ the ansatz for the right eigenvector of the boundary mode, 
 $|\psi_{\tilde{\mathbf{k}}}\rangle=\sum_{i=1}^{N_j}\beta^i|i\rangle\otimes |\xi_{\tilde{\mathbf{k}}}\rangle$, where 
$ |\xi_{\tilde{\mathbf{k}}}\rangle$ is a spinor, $i$ labels the latice sites along the $j$th direction, and $\beta$ is a scalar with $|\beta|<1$
(see SM~\cite{Supp} for details). 
By solving this Schr\"odinger equation we find that $\beta=M-\sum_{i}\cos \tilde{k}_i+i\lambda$
and that the boundary mode is an eigenstate of 
$i\Gamma_{d+1}\Gamma_j$ with eigenvalue $+1$. Hence, 
the projector $P$ onto the boundary space is given by $P=(1+i\Gamma_{d+1}\Gamma_j)/2$ and 
the effective boundary Hamiltonian is obtained by ${H}_b(\tilde{\mathbf{k}})=P({H}_0(\tilde{\mathbf{k}})+i\lambda \Gamma_{d+1})P$, with 
$\tilde{\mathbf{k}}$ satisfying $|\beta|=|M-\sum_{i}\cos \tilde{k}_i+i\lambda|<1$. 
Since $\Gamma_{d+1}$ anti-commutes with $\Gamma_j$, the non-Hermitian perturbation $i\lambda \Gamma_{d+1}$ vanishes under the projection. Thus, the effective boundary  Hamiltonian becomes ${H}_b(\tilde{\mathbf{k}})=\sum_{i\neq j} \sin k_i \gamma^i$, with $\gamma^i=P \Gamma_iP$, whose spectrum is manifestly real.
We note that while the effective boundary Hamiltonian is not altered by the non-Hermitian mass term $i\lambda \Gamma_{d+1}$, the range of $\mathbf{\tilde{k}}$ in which the boundary modes exists is changed to $|M-\sum_{i}\cos \tilde{k}_i+i\lambda|<1$.
 
To illustrate these general considerations, we consider as an example the two-dimensional topological insulator  
with a non-Hermitian mass term, i.e., ${H}_{\text{TI}}(\mathbf{k})=\sin k_x\Gamma_1+\sin k_y\Gamma_2
+(M-\cos k_x-\cos k_y)\Gamma_3+i\lambda\Gamma_3$.
As shown in Fig.~\ref{mFig2}, the spectrum of this non-Hermitian Hamiltonian is complex both with open and periodic boundary conditions.
With OBCs there appear  surface states within the region $|M-\cos k_x+i\lambda|<1$, whose spectrum is purely real.

\textit{Discussion.---}
In summary, we systematically investigated $d$-dimensional massive Dirac models on periodic lattices
perturbed by general non-Hermitian terms.
We find that there are three different types of non-Hermitian terms:
(i) non-Hermitian anti-commuting terms, (ii) non-Hermitian kinetic terms,
and (iii) non-Hermitian mass terms.
Interestingly, we find a two-fold duality for the first two types of non-Hermitian perturbations:
With open boundary conditions non-Hermitian anti-commuting terms give rise to intrinsic non-Hermiticity, while non-Hermitian kinetic terms lead to superficial non-Hermiticity. Vice versa, with periodic boundary conditions non-Hermitian anti-commuting terms induce superficial non-Hermiticity, while non-Hermitian kinetic terms generate intrinsic non-Hermiticity. Importantly, for the anti-commuting and kinetic terms the intrinsic non-Hermiticity manifests itself by exceptional points in the surface and bulk band structures, respectively. Non-Hermitian mass terms, in contrast, render the Hamiltonian always intrinsically non-Hermitian, independent of the boundary condition, but do not induce exceptional points in the band structure.

Our findings can be used as guiding principles for the design of applications in, e.g., photonic cavity arrays. 
For example, our analysis shows that topological Dirac systems perturbed by non-Hermitian mass terms exhibit robust surface states with a purely real spectrum [see Fig.~\ref{mFig2}(c)]. In photonic cavity arrays, these surface states provide robust channels for light propagation, which are protected against perturbations and disorder. 
Importantly, this property can be exploited for the design of efficient laser systems that are immune to disorder. 
That is, optically pumping the boundary of the cavity array induces single-mode lasing in the surface states, with high slope efficiency.
This was recently demonstrated, in two-dimensional optical arrays of microresonators~\cite{bahari_top_laser_science_2017,harari_topological_2018,bandres_topological_2018}.
It follows from our analysis that these phenomena  occur in a broader class of
photonic band structures, namely in any topological Dirac system~\cite{chiu_classification_2016} perturbed by non-Hermitian mass terms. 
 
Another possible application are photonic devices that utilize the exceptional points formed by the surface states of Dirac systems with non-Hermitian 
anti-commuting terms [see Fig.~\ref{mFig1}(b)]. While in non-periodic systems the creation of exceptional points
requires fine tuning, the surface exceptional points of the discussed lattice Dirac systems are required to exist by topology.
Moreover, these surface exceptional points are protected against
noise and disorder and cannot be removed by symmetry preserving perturbations (see SM~\cite{Supp} for details).
They could potentially be used for surface sensing.

We close by discussing several interesting directions for future studies. First, our approach can be generalized in a straightforward manner
to non-periodic Dirac models and to periodic gapless Dirac models, i.e., to non-Hermitian periodic lattices with semi-metallic band structures.
Second, it would be interesting to study the role of symmetries, specifically, how
the symmetries constrain the form of the non-Hermitian terms. 
Third, it would be of great value to derive a general and exhaustive classification of exceptional points based
on symmetry and topology~\cite{Budich_symmetry_2019}. This is particularly important in view of the numerous applications
of exceptional points in photonic devices. 
%This important fact deserves further investigations, both from a fundamental and applications point of view.
%can be extended to non-periodic systems, see SM~\cite{Supp}. 

\textit{Acknowledgments.---}
This work is supported by the NSFC (Grant No. 11874201), the Fundamental Research Funds for the Central Universities (Grant No. 0204/14380119), and the GRF of Hong Kong (Grant No. HKU173057/17P).

\bibliographystyle{apsrev4-1}

\bibliography{non-Hermi-topo}

 \widetext
 \clearpage
 

\section*{Supplementary material}

# Supplemental Material

**Authors:** W. B. Rui, Y. X. Zhao, and A. P. Schnyder

In this Supplemental Material, we solve the lattice model of the two-dimensional topological insulator with non-Hermitian anti-commuting terms considering both periodic and open boundary conditions and discuss the robustness of the exceptional points on the surface against symmetry preserving perturbations (Sec. I), derive the similarity transformation for the generic Dirac model with non-Hermitian kinetic terms for open boundary conditions (Sec. II), and calculate the boundary states for the generic Dirac model with non-Hermitian mass terms (Sec. III).

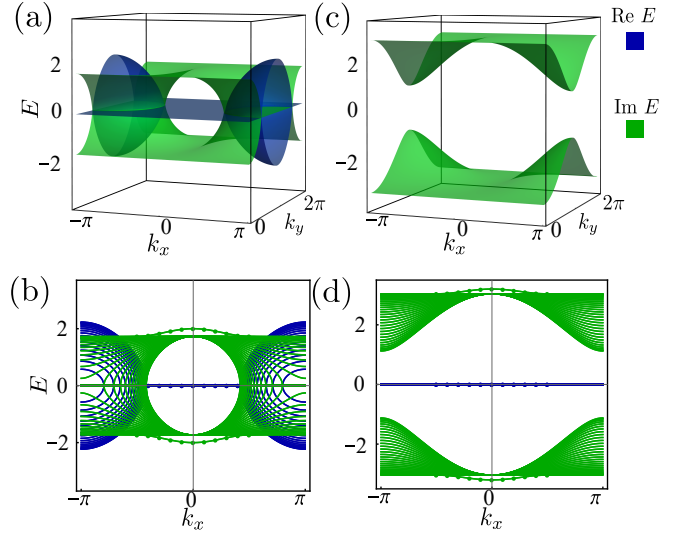

FIG. S1. The spectra for non-Hermitian 2D topological insulator with large non-Hermitian anti-commuting potentials. For $\min d_{\text{TI}}^2(\mathbf{k}) < \lambda^2 < \max d_{\text{TI}}^2(\mathbf{k})$, the bulk spectrum (a) is complex, and the corresponding open boundary spectrum is shown in (b); For $\max d_{\text{TI}}^2(\mathbf{k}) < \lambda^2$, the bulk spectrum (c) is purely imaginary, and the corresponding open boundary spectrum is plotted in (d). The dotted lines represent boundary states. The parameter $M$ is chosen to be 1.0.

## I. 2D TOPOLOGICAL INSULATOR WITH NON-HERMITIAN ANTI-COMMUTING TERMS

We start with the Hermitian lattice model of a 2D topological insulator,

$$\mathcal{H}_{\text{TI},0}(\mathbf{k}) = \sin k_x \Gamma_1 + \sin k_y \Gamma_2 + (M - \cos k_x - \cos k_y)\Gamma_3, \quad (S1)$$

where the five gamma matrices are $\Gamma_i = \sigma_i \otimes \tau_1$, $\Gamma_4 = \sigma_0 \otimes \tau_3$, $\Gamma_5 = \sigma_0 \otimes \tau_2$ ($i = 1, 2, 3$), with $\sigma_i$ and $\tau_i$ denoting the Pauli matrices. The energy eigenvalues are $E_{\text{TI},0}(\mathbf{k}) = \pm d_{\text{TI}}(\mathbf{k})$, with $d_{\text{TI}}^2(\mathbf{k}) = \sin^2 k_x + \sin^2 k_y + (M - \cos k_x - \cos k_y)^2$. The eigenstates are found to be

$$|+, \uparrow\rangle = \frac{1}{\sqrt{2}d_{\text{TI}}} \left(\sin k_x - i\sin k_y, \ -M(\mathbf{k}), \ 0, \ d_{\text{TI}}\right)^T,$$
$$|+, \downarrow\rangle = \frac{1}{\sqrt{2}d_{\text{TI}}} \left(M(\mathbf{k}), \ \sin k_x + i\sin k_y, \ d_{\text{TI}}, \ 0\right)^T,$$
$$|-, \uparrow\rangle = \frac{1}{\sqrt{2}d_{\text{TI}}} \left(-\sin k_x + i\sin k_y, \ M(\mathbf{k}), \ 0, \ d_{\text{TI}}\right)^T,$$
$$|-, \downarrow\rangle = \frac{1}{\sqrt{2}d_{\text{TI}}} \left(-M(\mathbf{k}), \ -(\sin k_x + i\sin k_y), \ d_{\text{TI}}, \ 0\right)^T, \quad (S2)$$

with $M(\mathbf{k}) = M - \cos k_x - \cos k_y$ and the matrix transposition T. The system is topologically non-trivial for $M \in (-2, 2)$.

Including the non-Hermitian anti-commuting perturbation $i\lambda\Gamma_4$, the Hamiltonian becomes

$$\mathcal{H}_{\text{TI}}(\mathbf{k}) = \mathcal{H}_{\text{TI},0}(\mathbf{k}) + i\lambda\Gamma_4. \quad (S3)$$

According to our theory, the non-Hermitian perturbation scatters eigenstates $\psi_0$ to $\Gamma_4\psi_0$. The energy eigenvalues become $E_{\text{TI}}(\mathbf{k}) = \pm\sqrt{d_{\text{TI}}^2(\mathbf{k}) - \lambda^2}$, and the corresponding eigenstates are modified as $\psi_\pm = \psi_0 + (iE_{\text{TI},0}/\lambda \pm \sqrt{1 - E_{\text{TI},0}^2/\lambda^2})\Gamma_4\psi_0$, with $\psi_0$ given in Eq. (S2).

For $d_{\text{TI}}^2(\mathbf{k}) > \lambda^2$, the spectrum is purely real in momentum space. According to the similarity transformation constructed below Eq. (4) in the main text, the Hamiltonian in Eq. (S3) can be converted to be Hermitian by the transformation

$$\mathcal{V}(\mathbf{k})^{-1}\mathcal{H}_{\text{TI}}(\mathbf{k})\mathcal{V}(\mathbf{k}) = \sqrt{1 - \frac{\lambda^2}{d_{\text{TI}}^2(\mathbf{k})}}\mathcal{H}_{\text{TI},0}(\mathbf{k}), \quad (S4)$$

where $\mathcal{V}(\mathbf{k}) = \exp[-\frac{i}{2}\mathcal{H}_{\text{TI},0}(\mathbf{k})\Gamma_4\eta_{\text{TI}}(\mathbf{k})/d_{\text{TI}}(\mathbf{k})]$, with $e^{\eta_{\text{TI}}(\mathbf{k})} = \sqrt{(d_{\text{TI}}(\mathbf{k}) + \lambda)/(d_{\text{TI}}(\mathbf{k}) - \lambda)}$. The matrix form of $\mathcal{V}(\mathbf{k})$ reads,

$$\mathcal{V}(\mathbf{k}) = \cosh\frac{\eta_{\text{TI}}(\mathbf{k})}{2} - i\sinh\frac{\eta_{\text{TI}}(\mathbf{k})}{2}\mathcal{H}_{\text{TI},0}(\mathbf{k})\Gamma_4/d_{\text{TI}}(\mathbf{k}). \quad (S5)$$

Taking open boundary conditions in the **y** direction with $N_y$ layers, the real space Hamiltonian reads

$$H_{\text{TI}}(k_x) = \frac{1}{2i}\left(\widehat{S} - \widehat{S}^\dagger\right) \otimes \Gamma_2 - \frac{1}{2}\left(\widehat{S} + \widehat{S}^\dagger\right) \otimes \Gamma_3$$
$$+ \mathbb{I}_{N_y} \otimes (\sin k_x \Gamma_1 + (M - \cos k_x)\Gamma_3) + \mathbb{I}_{N_y} \otimes i\lambda\Gamma_4. \quad (S6)$$

Here $\widehat{S}$ and $\widehat{S}^\dagger$ are the forward and backward translation operators in the **y** direction, which let $\widehat{S}|i\rangle = |i+1\rangle$ and $\widehat{S}^\dagger|i\rangle = |i-1\rangle$, where $i$ labels the $i$th site in the **y** direction. The corresponding matrix representations of

$\hat{S}$ and $\hat{S}^\dagger$ read

$$\hat{S} = \begin{pmatrix} 0 & 0 & 0 & 0 & \cdots & 0 \\ 1 & 0 & 0 & 0 & \cdots & 0 \\ 0 & 1 & 0 & 0 & \cdots & 0 \\ 0 & 0 & 1 & 0 & \cdots & 0 \\ \vdots & \vdots & \vdots & \ddots & \ddots & \vdots \\ 0 & 0 & 0 & 0 & 1 & 0 \end{pmatrix}, \hat{S}^\dagger = \begin{pmatrix} 0 & 1 & 0 & 0 & \cdots & 0 \\ 0 & 0 & 1 & 0 & \cdots & 0 \\ 0 & 0 & 0 & 1 & \cdots & 0 \\ 0 & 0 & 0 & 0 & \ddots & 0 \\ \vdots & \vdots & \vdots & \vdots & \ddots & 1 \\ 0 & 0 & 0 & 0 & 0 & 0 \end{pmatrix},$$
(S7)

with the dimension $N_y$.

Though the translational symmetry is broken in the **y** direction, it is still preserved in the **x** direction. We adopt the ansatz $|\psi_{k_x}\rangle = \sum_{i=1}^{N_y} \beta^i |i\rangle \otimes |\xi_{k_x}\rangle$ with $|\beta| < 1$ for the boundary states. Solving the Schrödinger equation $H_{\text{TI}}(k_x)|\psi_{k_x}\rangle = \hat{E}_{k_x}|\psi_{k_x}\rangle$ leads to the relation

$$\left[\sin k_x \Gamma_1 + \frac{1}{2i}(\beta - \beta^{-1})\Gamma_2 + (M - \cos k_x \right.$$
$$\left. - \frac{1}{2}(\beta + \beta^{-1}))\Gamma_3 + i\lambda\Gamma_4\right]|\xi_{k_x}\rangle = \hat{E}_{k_x}|\xi_{k_x}\rangle,$$
(S8)

in the bulk and

$$\left[\sin k_x \Gamma_1 + \frac{1}{2i}\beta\Gamma_2 + (M - \cos k_x - \frac{1}{2}\beta)\Gamma_3 \right.$$
$$\left. + i\lambda\Gamma_4\right]|\xi_{k_x}\rangle = \hat{E}_{k_x}|\xi_{k_x}\rangle,$$
(S9)

at the boundary. Taking the difference between Eqs. (S8) and (S9), we obtain a simpler constraint

$$i\Gamma_3\Gamma_2|\xi_{k_x}\rangle = |\xi_{k_x}\rangle,$$
(S10)

which means the boundary states correspond to the positive eigenvalue of $i\Gamma_3\Gamma_2$, from which we can construct the projector

$$P = \frac{1}{2}(1 + i\Gamma_3\Gamma_2).$$
(S11)

Applying this projector to Eq. (S9), we have

$$(\sin k_x \Gamma_1 + i\lambda\Gamma_4)|\xi_{k_x}\rangle = \hat{E}_{k_x}|\xi_{k_x}\rangle.$$
(S12)

With the relation of Eq. (S10), Eq. (S9) becomes $\left[\sin k_x \Gamma_1 + (M - \cos k_x - \beta)\Gamma_3 + i\lambda\Gamma_4\right]|\xi_{k_x}\rangle = \hat{E}_{k_x}|\xi_{k_x}\rangle$, and its difference with Eq. (S12) yields

$$\beta = M - \cos k_x.$$
(S13)

The effective boundary Hamiltonian can be obtained after the projection,

$$\mathcal{H}_{\text{TI,b}}(k_x) = P\mathcal{H}_{\text{TI}}(\mathbf{k})P = \sin k_x \gamma^1 + i\lambda\gamma^4,$$
(S14)

with $\gamma^1 = P\Gamma_1 P$ and $\gamma^4 = P\Gamma_4 P$, for $k_x$ satisfying $|\beta| = |M - \cos k_x| < 1$. The boundary energy eigenvalues are $E_{\text{TI,b}}(k_x) = \pm\sqrt{\sin^2 k_x - \lambda^2}$. For $|\lambda| < 1$, exceptional points emerge at $k_x = \pm \arcsin |\lambda|$.

In the main text, the non-Hermitian potential is relatively small and thus can be treated as perturbation. In FIG. S1, we increase the strength of the non-Hermitian potential. By increasing $\lambda$, the bulk system becomes gapless with complex spectrum (a), and gapless with purely imaginaly spectrum (c). The corresponding boundary states are denoted by dotted lines in (b) and (d), which exist even when the bulk becomes gapless.

### A. Robustness of the exceptional points on the surface against perturbations

In this subsection, we discuss the robustness of the surface exceptional points against symmetry preserving perturbations. Before going to specific perturbations, it is necessary to investigate the projection onto the surface subspace formed by the states of $\Psi = \{\Psi_\uparrow, \Psi_\downarrow\}$ corresponding to the positive eigenvalue of $i\Gamma_3\Gamma_2$ in Eq. (S10). It is found that after the projection, $\langle\Psi|\Gamma_1|\Psi\rangle = \sigma_1$, $\langle\Psi|\Gamma_4|\Psi\rangle = -\sigma_3$, and $\langle\Psi|\Gamma_5|\Psi\rangle = -\sigma_2$. Thus, the effective boundary Hamiltonian in Eq. (S14) is equivalent to $\mathcal{H}_{\text{TI,b}}(k_x) = \sin k_x \sigma_1 - i\lambda\sigma_3$ in the subspace spanned by $\Psi$.

The bulk Hamiltonian in Eq. (S3) is invariant under the time-reversal symmetry, i.e., $\hat{\mathcal{T}}^{-1}\mathcal{H}_{\text{TI}}(-\mathbf{k})\hat{\mathcal{T}} = \mathcal{H}_{\text{TI}}(\mathbf{k})$. Here the time-reversal operator is $\hat{\mathcal{T}} = \sigma_1 \otimes \tau_2 \hat{\mathcal{K}}$ with $\hat{\mathcal{K}}$ the complex conjugation. It is evident that $i\epsilon\Gamma_1$, $i\epsilon\Gamma_4$, and $i\epsilon\Gamma_5$ are perturbations that preserve time-reversal symmetry. Here $\epsilon$ is real and relatively small. In addition to time-reversal symmetry, the space-inversion symmetry is present if $\hat{\mathcal{P}}^{-1}\mathcal{H}_{\text{TI}}(-\mathbf{k})\hat{\mathcal{P}} = \mathcal{H}_{\text{TI}}(\mathbf{k})$ with $\hat{\mathcal{P}} = i\sigma_3 \otimes \tau_0$. The perturbations that are invariant under both time-reversal and space-inversion symmetry are $i\epsilon\Gamma_4$ and $i\epsilon\Gamma_5$. Since $i\epsilon\Gamma_4$ can be absorbed into the non-Hermitian potential $i\lambda\Gamma_4$, we focus on the perturbation $i\epsilon\Gamma_5$. After the inclusion of this perturbation, the effective boundary Hamiltonian becomes $\mathcal{H}_{\text{TI,b}}(k_x)' = \sin k_x \sigma_1 - i\lambda\sigma_3 - i\epsilon\sigma_2$ and the surface eigenvalues are $E' = \pm(\sin^2 k_x - \lambda^2 - \epsilon^2)$. The existence of the exceptional points is not affected, but their locations are shifted from $\pm \arcsin |\lambda|$ to $\pm \arcsin |\sqrt{\lambda^2 + \epsilon^2}|$.

## II. SIMILARITY TRANSFORMATION FOR THE GENERIC DIRAC MODEL WITH NON-HERMITIAN KINETIC TERMS FOR OPEN BOUNDARY CONDITIONS

With OBCs in the **y** direction the lattice Dirac model with a non-Hermitian kinetic perturbation reads

$$H(\tilde{\mathbf{k}}) = \frac{1}{2i}(\hat{S} - \hat{S}^\dagger) \otimes \Gamma_j - \frac{1}{2}(\hat{S} + \hat{S}^\dagger) \otimes \Gamma_{d+1}$$
$$+ \mathbb{I}_{N_j} \otimes (\sum_{i \neq j} \sin k_i \Gamma_i + (M - \sum_{i \neq j} \cos k_i)\Gamma_{d+1} + i\lambda\Gamma_j),$$
(S15)

which is superficially non-Hermitian. Note that for the part $(M - \sum_{i \neq j} \cos k_i)\Gamma_{d+1} + i\lambda\Gamma_j$ in the above Hamiltonian, the non-Hermitian term $i\lambda\Gamma_j$ can be regarded as a non-Hermitian anti-commuting perturbation. Hence, we can first make this part Hermitian by the following similarity transformation,

$$\rho_i^{-1}[(M - \sum_{i \neq j} \cos k_i)\Gamma_{d+1} + i\lambda\Gamma_j]\rho_i = \sqrt{M_k^2 - \lambda^2}\Gamma_{d+1}, \quad \text{(S16)}$$

with $M_k = M - \sum_{i \neq j} \cos M_k$. Here $\rho_i = (1+\alpha)\mathbb{I} + i(1-\alpha)\Gamma_j\Gamma_{d+1}$, with $\alpha = \sqrt{(M_k - \lambda)/(M_k + \lambda)}$. Its inverse is $\rho_i^{-1} = \frac{1}{4\alpha}[(1+\alpha)\mathbb{I} - i(1-\alpha)\Gamma_j\Gamma_{d+1}]$. Next we turn to the remaining terms with non-trivial spatial parts in Eq. (S15), which read

$$\frac{1}{2i}(\widehat{S} - \widehat{S}^\dagger) \otimes \Gamma_j - \frac{1}{2}(\widehat{S} + \widehat{S}^\dagger) \otimes \Gamma_{d+1}$$
$$= \widehat{S} \otimes (\frac{1}{2i}\Gamma_j - \frac{1}{2}\Gamma_{d+1}) - \widehat{S}^\dagger \otimes (\frac{1}{2i}\Gamma_j + \frac{1}{2}\Gamma_{d+1}). \quad \text{(S17)}$$

The operator $\rho_i$ acts on the internal degrees of freedom of the terms in the above equation as

$$\widehat{S} \otimes \rho_i^{-1}(\frac{1}{2i}\Gamma_j - \frac{1}{2}\Gamma_{d+1})\rho_i = \alpha\hat{S} \otimes (\frac{1}{2i}\Gamma_j - \frac{1}{2}\Gamma_{d+1}), \quad \text{(S18)}$$

$$\hat{S}^\dagger \otimes \rho_i^{-1}(\frac{1}{2i}\Gamma_j + \frac{1}{2}\Gamma_{d+1})\rho_i = \frac{1}{\alpha}\hat{S}^\dagger \otimes (\frac{1}{2i}\Gamma_j + \frac{1}{2}\Gamma_{d+1}). \quad \text{(S19)}$$

We construct the similarity transformation $\rho_S = \text{diag}(1, \alpha, \alpha^2, \cdots, \alpha^{N_j - 1})$ for the spatial part, which enables the following transformation

$$\rho_S^{-1}(\alpha\hat{S})\rho_S = \widehat{S}, \quad \text{(S20)}$$

$$\rho_S^{-1}(\frac{1}{\alpha}\widehat{S}^\dagger)\rho_S = \widehat{S}^\dagger. \quad \text{(S21)}$$

Finally, the full similarity transformation operator is constructed as

$$V = \rho_S \otimes \rho_i = \text{diag}(1, \alpha, \alpha^2, \cdots, \alpha^{N_j-1})$$
$$\otimes [(1+\alpha)\mathbb{I} + i(1-\alpha)\Gamma_j\Gamma_{d+1}], \quad \text{(S22)}$$

which converts the non-Hermitian Hamiltonian of Eq. (S15) into

$$V^{-1}H(\tilde{\mathbf{k}})V = \frac{1}{2i}(\widehat{S} - \widehat{S}^\dagger) \otimes \Gamma_j - \frac{1}{2}(\widehat{S} + \widehat{S}^\dagger) \otimes \Gamma_{d+1}$$
$$+ \mathbb{I}_{N_j} \otimes (\sum_{i \neq j} \sin k_i \Gamma_i + \sqrt{M_k^2 - \lambda^2}\Gamma_{d+1}), \quad \text{(S23)}$$

with $M_k = M - \sum_{i \neq j} \cos M_k$.

## III. BOUNDARY STATES FOR THE DIRAC MODEL WITH NON-HERMITIAN MASS TERMS

With PBCs the generic lattice Dirac model with non-Hermitian mass perturbations is given by

$$\mathcal{H}(\mathbf{k}) = \sum_{i=1}^d \sin k_i \Gamma_i + (M - \sum_{i=1}^d \cos k_i)\Gamma_{d+1} + i\lambda\Gamma_{d+1}. \quad \text{(S24)}$$

The energy eigenvalues are $E(\mathbf{k}) = \pm\sqrt{\sum_{i=1}^d \sin^2 k_i + (M + i\lambda - \sum_{i=1}^d \cos k_i)^2}$, which are complex in general.

With OBCs in the $\mathbf{j}$ direction, the real space Hamiltonian reads,

$$H(\tilde{\mathbf{k}}) = \frac{1}{2i}(\widehat{S} - \widehat{S}^\dagger) \otimes \Gamma_j - \frac{1}{2}(\widehat{S} + \widehat{S}^\dagger) \otimes \Gamma_{d+1}$$
$$+ \mathbb{I}_{N_j} \otimes (\sum_{i \neq j} \sin k_i \Gamma_i + (M - \sum_{i \neq j} \cos k_i)\Gamma_{d+1}) + \mathbb{I}_{N_j} \otimes i\lambda\Gamma_{d+1}. \quad \text{(S25)}$$

Here $N_j$ is the number of unit cells in the $\mathbf{j}$ direction. The spectrum of this Hamiltonian is also complex. For both Hamiltonians in Eqs. (S24) and (S25), which obey different boundary conditions, adding the non-Hermitian mass term is equivalent to making the Hermitian mass term complex, $M \to M + i\lambda$.

Except for the $\mathbf{j}$ direction, the translational symmetry in other directions is preserved. We adopt the ansatz $|\psi_{\tilde{\mathbf{k}}}\rangle = \sum_{i=1}^{N_j} \beta^i |i\rangle \otimes |\xi_{\tilde{\mathbf{k}}}\rangle$ with $|\beta| < 1$ for the boundary states. By solving the Schödinger equation $H(\tilde{\mathbf{k}})|\psi_{\tilde{\mathbf{k}}}\rangle = \hat{E}_{\tilde{\mathbf{k}}}|\psi_{\tilde{\mathbf{k}}}\rangle$ we find it gives two constraints,

$$[\sum_{i \neq j} \sin k_i \Gamma_i + \frac{1}{2i}(\beta - \beta^{-1})\Gamma_j + (M - \sum_{i \neq j} \cos k_i$$
$$- \frac{1}{2}(\beta + \beta^{-1}))\Gamma_{d+1} + i\lambda\Gamma_{d+1}]|\xi_{\tilde{\mathbf{k}}}\rangle = \hat{E}_{\tilde{\mathbf{k}}}|\xi_{\tilde{\mathbf{k}}}\rangle, \quad \text{(S26)}$$

and

$$[\sum_{i \neq j} \sin k_i \Gamma_i + \frac{1}{2i}\beta\Gamma_j + (M - \sum_{i \neq j} \cos k_i - \frac{1}{2}\beta)\Gamma_{d+1}$$
$$+ i\lambda\Gamma_{d+1}]|\xi_{\tilde{\mathbf{k}}}\rangle = \hat{E}_{\tilde{\mathbf{k}}}|\xi_{\tilde{\mathbf{k}}}\rangle. \quad \text{(S27)}$$

The difference between the above two equations yields a simpler relation,

$$i\Gamma_{d+1}\Gamma_j|\xi_{\tilde{\mathbf{k}}}\rangle = |\xi_{\tilde{\mathbf{k}}}\rangle. \quad \text{(S28)}$$

This means the boundary states correspond to the positive eigenvalue of $i\Gamma_{d+1}\Gamma_j$, from which we can construct the projector

$$P = \frac{1}{2}(1 + i\Gamma_{d+1}\Gamma_j). \quad \text{(S29)}$$

Notice with the relation of Eq. (S28), Eq. (S27) becomes,

$$\left[\sum_{i\neq j}\sin k_i\Gamma_i+(M-\sum_{i\neq j}\cos k_i+i\lambda-\beta)\Gamma_{d+1}\right]|\xi_{\tilde{\mathbf{k}}}\rangle$$
$$=\hat{E}_{\tilde{\mathbf{k}}}|\xi_{\tilde{\mathbf{k}}}\rangle, \quad (S30)$$

and under the projection $P$, Eq. (S27) also becomes,

$$\sum_{i\neq j}\sin k_i\Gamma_i|\xi_{\tilde{\mathbf{k}}}\rangle=\hat{E}_{\tilde{\mathbf{k}}}|\xi_{\tilde{\mathbf{k}}}\rangle. \quad (S31)$$

The difference between the above two equations gives,

$$\beta=M-\sum_{i\neq j}\cos k_i+i\lambda. \quad (S32)$$

We now calculate the effective boundary Hamiltonian by projection $P$. Remarkably, since the non-Hermitian mass term $i\lambda\Gamma_{d+1}$ anti-commutes with $i\Gamma_{d+1}\Gamma_j$ in the projector $P$, it will vanish after the projection. Thus, the resultant effective boundary Hamiltonian is

$$\mathcal{H}_b(\tilde{\mathbf{k}})=\sum_{i\neq j}\sin k_i\gamma^i, \quad (S33)$$

with $\gamma^i=P\Gamma_iP$, for $\tilde{\mathbf{k}}$ satisfying $|\beta|=|M-\sum_{i\neq j}\cos k_i+i\lambda|<1$. The localization region obtained in this way is identical to that obtained by the method of biorthogonal bulk-boundary correspondence [S1]. For the effective boundary Hamiltonian the boundary energy eigenvalues are purely real, i.e., $E_b(\tilde{\mathbf{k}})=\pm\sqrt{\sum_{i\neq j}\sin^2 k_i}$.

---



 \end{document}